\documentstyle[preprint,aps]{revtex}
\begin{document}
\title{ Induced Lorentz-Violating Chern-Simons Term in QED and
Anomalous Contributions to Effective Action Expansions }
\author{Lai-Him Chan}
\address{Department of Physics and Astronomy,
       	Louisiana State University,
	       Baton Rouge, Louisiana 70803-4001}
\date{ }
\draft
\maketitle

\begin{abstract}               
I present a unified formulation of  anomalous contributions in
quantum field theories by calculating directly the effective
action using the background field  and covariant-derivative
expansion technique. I use this method to determine uniquely the
induced Chern-Simons term from the a Lorentz and CPT violating
term in fermion QED Lagrangian. The outstanding ambiguity is
resolved by properly taken into account the noncommutivity of $A$
and $\partial$. The resulting vacuum polarization tensor acquires
a non-Feynman diagram anomalous contribution which accounts for
the discrepancy between the present calculation and the other
calculations.
\end{abstract}
\narrowtext
\def\sl#1{\rlap/#1}
\tightenlines

Recently there has been increasing interest in the possibility of
breaking Lorentz symmetry,  by adding Lorentz-violating terms to the
Lagrangian in the electromagnetic sector as well as the fermion
sector
\cite{cfj,ck,ck1,cg,cg2,jk,Oh,Jk,ChenChungFG,PVChung}. Since no
actual Lorentz violation has ever been observed, the upper limits of
these Lorentz-violating coupling constants are severely restrained.
The recent developments center on exploiting the unconventional
physical consequences which may either enhance the probability of
detecting such violations, or may set more stringent limits on the
Lorentz-violating terms\cite{cfj,ck,ck1,cg,cg2}. 

The electromagnetic Lorentz and CPT violating
Chern-Simons term is ${1\over2}\epsilon_{\mu\alpha\beta\nu}k^\mu 
F^{\alpha\beta}A^\nu\ $ \cite{cfj,ck},  where $k_\mu$ is a
constant 4-vector. It has been suggested
\cite{cfj} that astrophysical tests can provide a sensitive
measure of the magnitude of this Lorentz-violating Chern-Simons
term through the rotation of the  plane of polarization  of
radiation over cosmological distances.  Data from distant galaxies
has been analyzed and reanalyzed for radiation from cosmological
radio  sources to a high degree of accuracy with no convincing
evidence that such an effect exists\cite{cfj,noevid}. 

The quantum electrodynamics (QED) Lagrangian of a single charged
fermion field can be extended by inclusion of a CPT-odd Lorentz
violating term\cite{ck},

\begin{equation}\label{Lb}
{\cal L}=\bar{\psi}\Big[i\sl{\partial}-m-\gamma_5 \sl{b}-
Q\,\sl{\!A(x)}\Big]\psi\ ,
\end{equation}

where  $b_\mu$ is a constant 4-vector and $Q$ is the charge of the
fermion.  The effect of this violation can only be detected
indirectly through precision measurements of CPT violation of
physical properties of the fermion. The present knowledge of the
upper limit on $b_\mu$ is far less stringent than on $k_\mu$.

The interesting question is whether a Lorentz-violating CPT-odd
term in the fermion sector can induce a finite Lorentz-violating
Chern-Simons term in the  electromagnetic field  reminiscent of
the $\pi^0 \to
\gamma\gamma$ chiral anomaly \cite{abj,ad}. In that case,
without any unnatural fine-tuning or special constraints on model
building, the stringent limit on
$k_\mu$ would also imply a severe upper limit on $b_\mu$ so that
the  search for TCP violating effects from other high precision
measurements may not be effective. Since such a Chern-Simons term is not
gauge invariant,
it may be logical to expect that a gauge invariant Lagrangian
cannot induce a gauge non-invariant term in perturbation
theory \cite{cg2}.  However, it has been suggested \cite{ck} that
anomalous momentum surface integral terms of the type responsible for the
existence of chiral anomaly may also contribute to a  finite value
of $k_\mu$. Theories with anomaly are theories with one-loop anomalous contribution to
the effective Lagrangian which violate the original gauge symmetry.
While the recognition of anomalous contributions in
quantum field theories has led to tremendous progress in particle
physics, it remains a puzzle why quantum field theory can allow such
undeterministic ambiguity\cite{Jk}. In this paper, I propose how this puzzle
of ambiguous anomalous contributions can be resolved. The solution may
provide  a necessary clue to the understanding the general problem of
anomalies in quantum field theories.

The effective action is obtained by integrating out the fermion field in
Eq.~\ref{Lb},

\begin{equation}\label{Actionb}
{\Gamma}_{eff}=-i\,Tr\,ln\Big[i\sl{\partial}-Q\,\sl{\!A(x)}-m 
-\gamma_5\sl{b}\Big]\ ,
\end{equation}
where $Tr$ stands for the trace in configuration space, spinor
matrix space and internal symmetry space if it is included. 
The trace in configuration space is carried out by resolving
$\delta^4(x-x')=\int {d^4 p\over(2\pi)^4}e^{-ip.(x-x')}$\cite{cs}, and the
effective Lagrangian is given by

\begin{equation}\label{Leffb}
{\cal L}_{eff}=-i\int\!\!{d^4
p\over(2\pi)^4}tr\,ln\Big[\sl{p}-m+i\sl{\partial}-Q\,\sl{\!A(x)}
-\gamma_5\sl{b}\Big].
\end{equation}

Here $p$ is the internal loop momentum and $\partial$ corresponds
to the routing momentum of the external lines when the effective
action is expanded into Feynman diagrams. It becomes important to
note that the trace ``$tr$" here does not apply to space-time. In 
any perturbation expansion, one partitions the argument of the
logarithmic function into two parts, $\sl{p}-m+X$ and
$Y$, and expands the effective action  using modified Hausdorf
expansion,

\begin{eqnarray*}
&tr&ln\Big[\sl{p}\!-\!m\!+\!X\!+\!Y\Big]=tr\,ln(\sl{p}\!-\!m\!+\!X)
(1+{1\over\sl{p}\!-\!m\!+\!X}Y)\nonumber\\ 
&=&tr\Big\{\!ln(\sl{p}\!-\!m\!+\!X)+ln(1+{1\over\sl{p}-m+X}Y)
+{1\over2}\big[ln(\sl{p}\!-\!m\!+\!X),
ln(1+{1\over\sl{p}\!-\!m\!+\!X}Y)\big]\\&+&{\rm higher\ order\
commutators}\Big\}.
\end{eqnarray*}

Irrespective of any gamma matrix dependence, if the space-time
parts of the operators $X$ and $Y$ do not commute, as in the case of a
Feynman graph expansion where 
$X=i\sl{\partial}+\dots$ and $Y$ is a function of $x$, the commutator terms
may not be zero. However, the loss of cyclic permutation invariance of the
space-time operators can be compensated by an appropriate shift in the momentum
integration variable $p$. As it is
well-known from the axial anomaly, a shift of the variable of integration for
a linear divergent integral contributes a nonzero finite change. For this
reason, the commutator terms give a well-defined additional anomalous
contribution to the effective Lagrangian beyond those calculated
from the Feynman diagram reconstruction. If these
commutators are taken into account, the results of any calculation should be
independent of how expansions are carried out.
 We can avoid calculations of these commutators if the initial
expansion is carried out by choosing $X$ and $Y$ so that the
contribution from  all commutator terms vanishes. A symmetrical
choice is $X=0$. For the effective action Eq. \ref{Actionb}  where
$b_\mu$ is constant, any combination keeping
$\Pi_\mu=i\partial_\mu-Q A_\mu$  as a single entity will
satisfy this condition.  They should all yield the same
unique result with no ambiguity. The essence of the
covariant-derivative-expansion is to develop a series of local effective
Lagrangian in powers of
$\Pi_\mu$, rather than in powers of $i\partial_\mu$ and $A_\mu$
separately\cite{cd,MaryK}. The choice
$X=i\sl{\Pi}$ and $Y=-\gamma_5\sl{b}$ gives the simplest calculation.
Expanding the effective action first in powers of $\gamma_5\sl{b}$, 
and then in powers of $i\sl{\Pi}$, we can calculate the Chern-Simons
term,

\begin{eqnarray}
&{\cal L}_{eff}&=i\int\!\!{d^4p\over(2\pi)^4}tr
{1\over\sl{p}-m+\sl{\Pi}}\gamma_5\sl{b}+\dots\nonumber
\\
 &=&-i\,\int\!\!{d^4p\over(2\pi)^4}
tr\,S\sl{\Pi}S\sl{\Pi}S\sl{\Pi}S\gamma_5\sl{b}+\dots\label{Leff1}\\
 &=& \!\!
4b^\mu \Pi^{\alpha}\Pi^\beta\Pi^\kappa
\epsilon_{\lambda\alpha\beta\nu}\!\!
\int\!\! {d^4p\over(2\pi)^4}\!\left[{g_\kappa^\nu \over (p^2-m^2)^2} 
-{4p_\kappa p^\nu \over (p^2-m^2)^3}\right]
\nonumber
\end{eqnarray}

where $S={i\over \sl{p}-m}$.   The
Lorentz and CPT violating Chern-Simons term is finite without any
regularization, $k_\mu={1\over8\pi^2}Q^2 b_\mu$.

We now consider the general case of a massive fermion
Lagrangian with $m\ne 0$,
\begin{equation}\label{Lfermiongeneral} {\cal L}
=\bar{\psi}\Big[i\sl{\partial}-m-Q \,\sl{\!A(x)}-\chi(x)\Big]\psi\ ,
\end{equation} 
where $\chi(x)$ can include any combination of boson fields and  any
dependence on gamma matrices. The corresponding effective Lagrangian
is 
\begin{eqnarray}
{\cal L}_{eff}=-i\!\int\!\!
{d^4p\over(2\pi)^4}tr\,\{ln\Big[\sl{p}+\sl{\Pi}\!-\!m\!-\!\chi(x)
\Big]-ln(\sl{p}-m)\}
\nonumber.\end{eqnarray}
To isolate the anomalous part from the normal vector gauge
invariant part of the effective action, I employ the
transformation of Ref.\cite{MaryK}, with the exception that
I do not drop those  momentum surface integral terms, which are
 the gauge-noninvariant contributions. I define the gauge invariant
combination to be\cite{MaryK}, 
\begin{eqnarray*}
T=&&e^{-\Pi\cdot{\partial\over\partial p}}\,tr\, ln\Big[\sl{p}
+\sl{\Pi}-m-\chi(x)\Big]e^{\Pi\cdot{\partial\over\partial
p}}\\
=&&tr\,ln\Biggl[
\sl{p}-m-{i\over 2}Q\gamma^\mu \sum_{n=0}^\infty{
2(n+1)\over(n+2)!} {1\over i^n}  (D_{\mu_1}\!.. D_{\mu_n}
\!F_{\mu\nu}){\partial^{\,n+1}\over\partial p_{\mu_1}\!..\partial
p_{\mu_n}\partial p_\nu}\nonumber\\
&&\quad-\sum_{n=1}^\infty{1\over n!} {1\over i^n}(D_{\mu_1}\!..
D_{\mu_n}\chi){\partial^{\,n}\over\partial p_{\mu_1}\!..\partial
p_{\mu_n}}-\chi(x)\Biggr],\label{Leffcovariant}
\end{eqnarray*}
where the covariant derivatives are 
defined by $D_\mu
\phi={1\over i}[\Pi_\mu,\phi]$ and $[\Pi_\mu,\Pi_\nu]=
[i\partial_\mu\!-\!Q A_\mu,i\partial_\nu\!-\!Q A_\nu]=-iQ F_{\mu\nu}$.
The gauge invariant part of the effective Lagrangian is,
\begin{equation}
{\cal L}_{eff}^{inv}=-i\,\int
\!\!{d^4p\over(2\pi)^4}\big[T-ln(\sl{p}-m)\big].
\end{equation}
The gauge-noninvariant anomalous part is,
\begin{eqnarray*}
{\cal L}_{eff}^{nonin}  
=-i\!\int \!\!{d^4 p\over(2\pi)^4}\Big(
e^{\Pi\cdot{\partial\over\partial
p}}-1 \Big)T=-i\!\int \!\!{d^4 p\over(2\pi)^4}\Big[
\sum_{n=1}^\infty {1\over n!}
\Big(\Pi\!\cdot\!{\partial\over\partial p}\Big)^{\!n} T\Big].
\end{eqnarray*}
 The summation term is the anomalous
effective Lagrangian defined in this paper as the vector
gauge noninvariant part. It would vanish completely if
conventional gauge invariant regularization such as Pauli-Villars
regularization or the dimensional regularization is used. The
noncovariant part is directly tied to a total momentum derivative.
Therefore the anomalous effective Lagrangian can be expressed as a
surface integral. If the unadulterated dimensional
regularization is used to evaluate the momentum integral, the
anomalous effective Lagrangian vanishes. Independently, it is also
clear that the power counting of the momentum of the integrand
becomes sufficiently negative as the number of differentiations
increases, so that the surface integral would vanish naturally  for
sufficiently high value of
$n$ in the sum, and only very few terms would contribute regardless
of regularization.

Conventionally, regularization procedures are used to keep track of 
degrees of divergence for the purpose of renormalization. For finite
terms, regularization should make no difference. However, for the
anomalous effective Lagrangian,  gauge invariant regularization
schemes force all terms in the anomalous effective Lagrangian to 
be zero. For those finite terms, the role of regularization is no
longer regulating divergent quantities but rather of imposing gauge
invariance. If we take the stance that processes not absolutely
forbidden can exist and  exact
symmetry can be broken by radiative corrections, we should allow the
possibility of not imposing gauge invariant regularization on
nondivergent terms unless ambiguity and inconsistency occur.

For the calculation of the Lorentz violating Chern-Simons term,
I choose $\chi=\gamma_5\sl{b}$ and select only the relevant
Chern-Simons terms,
\begin{eqnarray}
&{\cal L}_{eff}^{CS}  
&=-i\!\int \!\!\!{d^4
p\over(2\pi)^4}
\Pi\!\cdot\!\!{\partial\over\partial p}  tr\,ln\Big[
\sl{p}-m-\!{i\over 2}Q\gamma^\mu 
F_{\mu\nu}{\partial\over\partial
p_\nu}-\gamma_5\sl{b}\Big]\nonumber\\ 
&=&i\int \!\!\!{d^4
p\over(2\pi)^4}tr
\Pi\!\cdot\!\!{\partial\over\partial p} \Big[
{1\over \sl{p}-m}{i\over 2}Q\gamma^\mu 
F_{\mu\nu}{\partial\over\partial p_\nu}{1\over
\sl{p}-m}\gamma_5\sl{b}\Big]\nonumber\\
&=&\!\!{Q\over 2}\!\!\int \!\!\!{d^4 p\over(2\pi)^4}[\Pi_\alpha
F_{\mu\beta}+\!\Pi_\mu F_{\beta\alpha}+\!\Pi_\beta F_{\mu\alpha}]
  tr S\gamma^\alpha S\gamma^\beta S\gamma^\mu
S\gamma_5\sl{b}\nonumber\\  
&=&{1\over16\pi^2}Q^2
\epsilon_{\mu\alpha\beta\nu}b^\mu F^{\alpha\beta} A^\nu \ .
\end{eqnarray}
I recover the same Chern-Simons term. It also becomes obvious that
there is no anomalous contributions from higher order $b_\mu$, since
terms higher powers in $b$ are necessarily accompanied by more
convergent factor which renders the surface integrals
zero\cite{PVChung}.

Now I have traced that the  ambiguity of the anomalous
contribution is due to the inappropriate expansion of the ``tr\,ln"
function after the space-time trace has been evaluated. The problem
of extra contributions due to nonvanishing commutators would not be
present if the expansion can be performed before the space-time
trace is carried out. However due to short distant singularity,
we cannot carry out such an expansion directly on Eq.~\ref{Actionb}
without proper regularization. A very natural
regularization has indeed been proposed to simplify the calculation
of effective action covariant derivative expansion with internal
symmetry\cite{cd}.
The effective action in Eq.~\ref{Actionb} is invariant under a finite
momentum translation,
\begin{eqnarray}
{\Gamma}_{eff}&=&-i\,Tr\,e^{ip\cdot x} ln\Big[i\sl{\partial}-
Q\,\sl{\!A(x)}-m  -\gamma_5\sl{b}\Big] e^{-ip\cdot x}\nonumber\\&=&
-i\,Tr\,ln\Big[\sl{p}+
\sl{\Pi}-m  -\gamma_5\sl{b}\Big]\ .
\end{eqnarray}
The arbitrary momentum $p$ can then be averaged over
the entire momentum space,
\begin{eqnarray}
{\Gamma}_{eff}=-i{1\over\delta^4(0)}\int{d^4p\over(2\pi)^4}
\,Tr\,ln\Big[\sl{p}+\sl{\Pi}-m  -\gamma_5\sl{b}\Big]\ ,
\end{eqnarray}
where $\delta^4(0)=\int\!{d^4p\over(2\pi)^4}$. Only
after averaging can we fully use the cyclic permutation of the
full trace and  freely expand the ``$ln$" function without any
possible ambiguity due to extra commutator terms. The
Chern-Simons actions from any possible expansion will all end up
equivalent to the following form:
\begin{equation}
{\Gamma}_{eff}^{CS}
 =-i\,{1\over\delta^4(0)}\int\!\!{d^4p\over(2\pi)^4}
Tr\,S\sl{\Pi}S\sl{\Pi}S\sl{\Pi}S\gamma_5\sl{b}\ ,
\end{equation}
which is the counterpart of Eq.~\ref{Leff1}. The calculations of
taking trace of the gamma matrices and integrating the momentum are
the same as before. The corresponding part of the calculation
$\int\!\! d^4x\,\Pi^{\alpha}\Pi^\beta\Pi^\kappa
\epsilon_{\lambda\alpha\beta\nu}$ is 
\begin{eqnarray}
{1\over\delta^4(0)}&{\cal T}r&
\Pi^{\alpha}\Pi^\beta\Pi^\kappa\epsilon_{\lambda\alpha\beta\nu}=
\!\!\int\!\! d^4x{1\over\delta^4(0)}\langle
x|\Pi^{\alpha}\Pi^\beta\Pi^\kappa|x\rangle
\epsilon_{\lambda\alpha\beta\nu}\nonumber\\
&=&-{i\over2}Q\!\int\!\!d^4x{1\over\delta^4(0)}\langle
x|F^{\alpha\beta}(i\partial^\kappa-Q A^\kappa)|x\rangle
\epsilon_{\lambda\alpha\beta\nu}\nonumber\\
&=&\int\!\!d^4x{i\over2}Q^2
F^{\alpha\beta} A^\kappa
\epsilon_{\lambda\alpha\beta\nu}\ ,
\end{eqnarray}
where I have used the identities
$\langle x|i\partial^\kappa|x\rangle=0$ and
$\langle x|x\rangle=\delta^4(0)$.
Therefore I arrive at the same Chern-Simons effective Lagrangian
again.

The entanglement of expanding two noncommuting operators may also be
bypassed if we differentiate the expression with respect to $m$ first,
\begin{equation} {\partial \over \partial m}  
tr\,ln\Big[\sl{p}-m+X+Y\Big]=-tr{1\over \sl{p}-m+X+Y}
\end{equation} 
and then the identity ${1\over A+B}={1\over A}-{1\over A}B{1\over
A+B}$ may be repeatedly used for further expansion. The final result
should be independent of  how the expansion is carried
out. I rewrite the effective Lagrangian Eq.~\ref{Leffb}:
\begin{equation}
{\cal L}_{eff}=-i\,\int\! {d^4
p\over(2\pi)^4}\int_m^\infty\! dm\,tr\,{1\over\sl{p}-m+\sl{\Pi}
-\gamma_5\sl{b}}\label{meff}\ ,
\end{equation}
and expand it in powers of $\sl{\Pi}$ and $\gamma_5\sl{b}$ to obtain 
the Chern-Simons Lagrangian,
\begin{eqnarray}
&{\cal L}_{eff}^{CS}&=-\!\int\! \!\!{d^4
p\over(2\pi)^4}\!\int_m^\infty\!
\!\!dm\,tr\Big[S\sl{\Pi}S\sl{\Pi}S\sl{\Pi}S\gamma_5
\sl{b}S+ S\sl{\Pi}S\sl{\Pi}S\gamma_5\sl{b}S\sl{\Pi}S\nonumber \\
&&\qquad \qquad +\,
S\sl{\Pi}S\gamma_5\sl{b}S\sl{\Pi}S\sl{\Pi}S+
S\gamma_5\sl{b}S\sl{\Pi}S\sl{\Pi}S\sl{\Pi}S\Big]\\
&=&4i\,\epsilon_{\mu\alpha\beta\nu}b^\mu
\Pi^{\alpha}\Pi^{\beta}\Pi^\nu\!
\int\!\! \!{d^4
p\over(2\pi)^4}{m^2\over (p^2-m^2)^3}={i\over16\pi^2}Q^2
F^{\alpha\beta} A^\kappa
\epsilon_{\lambda\alpha\beta\nu}\nonumber.
\end{eqnarray} 
The perturabtion expansion in powers of $A$ can now be carried
out easily from the expression in Eq.\ref{meff} to yield the
quadratic term in $A$,
\begin{eqnarray}
&{\cal L}_{eff}^{A^2}&=i\,Q^2\!\!\int\!\! {d^4
p\over(2\pi)^4}\!\int_m^\infty\!\!\!dM\,tr\,
{1\over\sl{p}\!-\!M\!+\!i\sl{\partial}\!-\!\gamma_5\sl{b}}\,\, 
\sl{\!A}(x){1\over\sl{p}\!-\!M\!+\!i\sl{\partial}\!-\!\gamma_5\sl{b}}
\,\,\sl{\!A}(x){1\over\sl{p}\!-\!M\!+\!i\sl{\partial}\!-\!
\gamma_5\sl{b}}\\
&=&-\int\! {d^4 p\over(2\pi)^4}{i\over2}\,Q^2 A_\mu(p)A_\nu(-p)\Pi^{\mu\nu}(p).
\end{eqnarray}
The vacuum polarization tensor
\begin{eqnarray}
&\Pi^{\mu\nu}(p)&=\int\!
{d^4 k\over(2\pi)^4}tr\,\Bigg\{\gamma^\mu{i\over
\sl{k}-m-\gamma_5\sl{b}}\gamma^\nu{i\over
\sl{k}+\sl{p}-m-\gamma_5\sl{b}}\label{vacp}\\
&+&\!\!\!\!\!i\!\int^\infty_m\!\!dM\Big[\gamma^\mu
\Big({i\over\sl{k}\!-\!M\!-\!\gamma_5\sl{b}}\Big)^{\!\!2}\!
\gamma^\nu {i\over\sl{k}+\sl{p}\!-\!M\!-\!\gamma_5\sl{b}}
-\gamma^\mu\Big({i\over
\sl{k}-\sl{p}\!-\!M\!-\!\gamma_5\sl{b}}\Big)^{\!\!2}\!\gamma^\nu
{i\over\sl{k}\!-\!M\!-\!\gamma_5\sl{b}}\Big]\Bigg\}\nonumber
\end{eqnarray}
has been rewritten into an expression in terms of the standard
one-loop vacuum polarization Feynman amplitude with
$b_\mu$-exact fermion propagators
\cite{jk,Oh,PVChung} plus a $b_\mu$-exact correction from the
noncommutativity of the operators $A$ and $\partial$.  This
correction, even though expressed in a closed form via a mass
integration representation as the difference of two terms differed
by a shift of integration variable ($k\to k-p$), is nonetheless
nonzero. It gives precisely the additional contribution 
$-{1\over32\pi^2}Q^2b_\mu $ to $k_\mu$ to combine with the previous
calculation\cite{jk,Oh,PVChung} of the first term to yield
$k_\mu={3\over32\pi^2}Q^2b_\mu-{1\over32\pi^2}Q^2b_\mu={1\over8\pi^2}Q^2b_\mu$.
Therefore we have shown that there is no ambiquity in the
calculation of
$k_\mu$ as long as the  noncommutativity of the operators $A$ and
$\partial$ is taken into account. Whether the calculation is done
nonperturbatively in $b_\mu$ or not makes no difference. 

 The answer to the question whether a Lorentz-violating CPT-odd
term in the fermion sector can induce a finite Lorentz-violating
Chern-Simons term in the  electromagnetic field hinges on the
answer to a much more general question: must an apparently finite
nongauge invariant effective Lagrangian term from radiative
correction be forced to vanish by imposing a gauge invariant
regularization on it? The authors of Ref.\cite{jk} remind us that
an alternative must be allowed in order for the axial anomaly to
exist. However, within this new domain, the conventional wisdom
in reconstructing the effective Lagrangian from on the Feynman
diagram expansion becomes ambiguous. There is an intrinsic
quantum incompatibility between the power expansion of the photon
fields and the derivative (small momentum) expansion. The subtlety
requires a correction to the standard perturbation theory
expansion similar in spirit to the modification of the T-product
to the $T^*$-product. The calculation can be made much simpler by
using the back ground field  formulation of the effective action
covariant-derivative-expansion.  While the
covariant-derivative-expansion preserves gauge symmetry when the
symmetry is exact, it also serves as a regulator when  gauge
symmetry is forced to be broken by radiative corrections. 
 I have used different approaches to explore various aspects of
the anomalous contribution to the effective Lagrangian. The
calculations from various  covariant derivative expansions all
converge to
 a unique Chern-Simons term
${1\over16\pi^2}Q^2\epsilon_{\mu\alpha\beta
\nu}b^\mu F^{\alpha\beta} A^\nu $. I have also derived the
complete expression for the vacuum polarization tensor to account
for the discrepancy between my calculation and the previous
calculations\cite{jk,Oh}. Since the effect of the Chern-Simons
term has not been detected, the minor difference in its numerical
coefficient may not appear to have much significance. However in
resolving the ambiguity in its calculation, I have demonstrated
how the modern background field calculation and the effective
action expansion method can be used to understand the complexity
of quantum field theory calculations. This new approach may shed
more light on the anomalous contributions to the quantum field
theories.

I thank Roman Jackiw, Alan Kosteleck\'{y}, J.-M. Chung and P. Oh
for sending me advance copies of their manuscripts and for 
helpful discussions.

\end{document}